\documentclass[aps,prb,showpacs,amsmath,twocolumn,amssymb,superscriptaddress,letterpaper]{revtex4}
\usepackage{graphicx}
\usepackage{dcolumn}
\usepackage{bm}
\usepackage{xcolor}

\newcommand{\bs}[1]{\boldsymbol{#1}}

\begin{document}
\renewcommand{\thefootnote}{\fnsymbol{footnote}}
\title{Ultraquantum magnetoresistance in Kramers Weyl semimetal candidate $\beta $-Ag$_2$Se
}

\author{Cheng-Long Zhang}
\affiliation{International Center for Quantum Materials and School of Physics, Peking University, Beijing 100871, China}

\author{
Frank Schindler}
\affiliation{
 Department of Physics, University of Zurich, Winterthurerstrasse 190, 8057 Zurich, Switzerland}

\author{Haiwen Liu}
\affiliation{ Center for Advanced Quantum Studies, Department of Physics, Beijing Normal University, Beijing, 100875, China}

\author{Tay-Rong Chang}
\affiliation{Department of Physics, National Tsing Hua University, Hsinchu 30013, Taiwan}

\author{Su-Yang Xu}
\affiliation{Joseph Henry Laboratory, Department of Physics, Princeton University, Princeton, New Jersey 08544, USA}

\author{Guoqing Chang}
\affiliation{Centre for Advanced 2D Materials and Graphene
Research Centre National University of Singapore,
6 Science Drive 2, Singapore 117546}
\affiliation{Department of Physics, National University of Singapore,
2 Science Drive 3, Singapore 117542}

\author{Wei Hua}
\affiliation{Department of Chemistry, Peking University, Beijing 100871, China}

\author{Hua Jiang}
\affiliation{College of Physics, Optoelectronics and Energy, Soochow University, Suzhou 215006, China}

\author{Zhujun Yuan}
\affiliation{International Center for Quantum Materials and School of Physics, Peking University, Beijing 100871, China}

\author{Junliang Sun}
\affiliation{Department of Chemistry, Peking University, Beijing 100871, China}

\author{Horng-Tay Jeng}
\affiliation{Department of Physics, National Tsing Hua University, Hsinchu 30013, Taiwan}
\affiliation{Institute of Physics, Academia Sinica, Taipei 11529, Taiwan}

\author{Hai-Zhou Lu}
\affiliation{Department of Physics, South University of
Science and Technology of China, Shenzhen, China}

\author{Hsin Lin}
\affiliation{Centre for Advanced 2D Materials and Graphene
Research Centre National University of Singapore,
6 Science Drive 2, Singapore 117546}
\affiliation{Department of Physics, National University of Singapore,
2 Science Drive 3, Singapore 117542}

\author{M. Zahid Hasan}
\affiliation{Joseph Henry Laboratory, Department of Physics, Princeton University, Princeton, New Jersey 08544, USA}

\author{X. C. Xie}
\affiliation{International Center for Quantum Materials and School of Physics, Peking University, Beijing 100871, China}
\affiliation{Collaborative Innovation Center of Quantum Matter, Beijing, China}

\author{
Titus Neupert}
\affiliation{
 Department of Physics, University of Zurich, Winterthurerstrasse 190, 8057 Zurich, Switzerland}

\author{Shuang Jia}
\email{gwljiashuang@pku.edu.cn}
\affiliation{International Center for Quantum Materials and School of Physics, Peking University, Beijing 100871, China}
\affiliation{Collaborative Innovation Center of Quantum Matter, Beijing, China}
\begin{abstract}
The topological semimetal $\beta $-Ag$_2$Se features a Kramers Weyl node at the origin in momentum space and a quadruplet of spinless Weyl nodes, which are annihilated by spin-orbit coupling. We show that single crystalline \nolinebreak{$\beta $-Ag$_2$Se} manifests giant Shubnikov-de Haas oscillations in the longitudinal magnetoresistance which stem from a small electron pocket that can be driven beyond the quantum limit by a field less than 9 T. This small electron pocket is a remainder of the spin-orbit annihilated Weyl nodes and thus encloses a Berry-phase structure. Moreover, we observed a negative longitudinal magnetoresistance when the magnetic field is beyond the quantum limit. Our experimental findings are complemented by thorough theoretical band structure analyses of this Kramers Weyl semimetal candidate, including first-principle calculations and an effective $\bs{k}\cdot\bs{p}$ model.

\end{abstract}
\pacs{72.15.Gd  71.70.Di  72.15.Lh }
\maketitle

\section{INTRODUCTION}
\begin{figure*}[t]
\centering
  \includegraphics[width=0.98\textwidth]{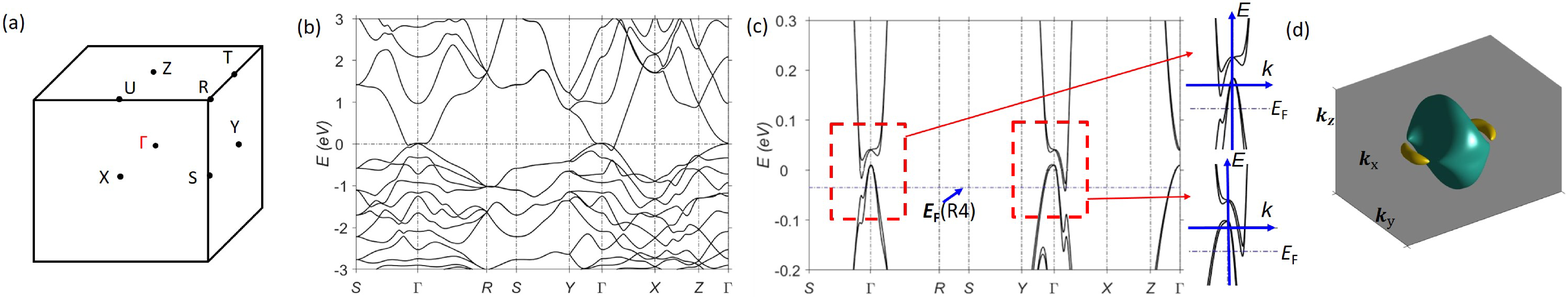}\\
  \caption{Band structure of $\beta $-Ag$_2$Se.
  (a) High-symmetry points in the Brillouin zone of space group $P2_12_12_1$.
  (b) Band structure without SOC. The band crossing along the $\Gamma$-X and $\Gamma$-Y lines are spinless Weyl points with opposite chiral charge.
 (c) Band structure with SOC. The spinless Weyl nodes annihilate each other when SOC is increased to the physical value. What remains of them are the characteristic extrema of the bands that form electron-like pockets along the $\Gamma$-X line. In addition, the lifting of the band degeneracy leads to the formation of Kramers Weyl points at each band crossing at the $\Gamma$ point.
 Two zoomed-in panels show details on the band structure at the $\Gamma$ point.
 (d) Fermi surface maps at energy $E_\mathrm{F}$ marked in panel (c). The green Fermi surface encloses a Kramers Weyl point, while the large SdH oscillations measured in this work are attributed to the yellow pockets.   }
  \label{Fig6}
\end{figure*}
Symmetry and topology can cooperate in electronic solids to create low-energy electronic structures with unique properties, resembling Weyl fermions, Dirac fermions, nodal line fermions, and many more \cite{Weyl_wanxiangang,TaAs_Chenyulin,TaAs_theory_Hsin,TaAs_theory_Xidai,TaAs_Arpes_Hongding,TaAs_Arpes_Hasan,nodal_burkov,bian2016topological,schoop2016dirac,Madhab_ZrSiS,pts_ZCL,bradlyn2016beyond,soluyanov2015type,wang2016hourglass}.
The two-fold band degeneracy associated with Weyl fermions is most fundamental, as it is protected by translation symmetry alone and can therefore appear at generic, low-symmetry positions in momentum space. This is indeed the case in all so-called band-inverted Weyl semimetals that have been confirmed experimentally to date.

It has been shown theoretically that there exists a second class of Weyl semimetals in which Weyl nodes are pinned to time-reversal symmetric momenta (TRIMs) in the Brillouin zone and cannot move freely in momentum space under small perturbations.~\cite{bradlyn2016beyond,chang2016kramers} These Kramers Weyl nodes appear generically in all chiral crystals, i.e., in crystals that have a sense of handedness by lacking any roto-inversion symmetries \cite{chang2016kramers}.
In time-reversal symmetric systems, Kramers theorem enforces two-fold band degeneracy at every TRIM. Spin-orbit coupling (SOC) can in principle split the two Kramers degenerate bands in every direction away from the TRIM, leaving behind a Weyl cone.
Roto-inversion and general nonsymmorphic symmetries may prevent the spin-orbit induced lifting of the band degeneracies along certain directions in momentum space, thereby preventing the Weyl cone formation. In symmorphic chiral crystals, however, every Kramers pair of bands at every TRIM is guaranteed to host a Weyl cone. For non-symmorphic chiral crystals, the latter is true for a subset of TRIMs only, which however always includes the $\Gamma$ point.

Here, we present a magnetotransport study combined with detailed analyses of the low-energy electronic structure of $\beta $-Ag$_2$Se (Naumannite), which is among the first materials that have been theoretically proposed to be a candidate Kramers Weyl semimetal. $\beta $-Ag$_2$Se crystallizes in a nonsymmorphic chiral structure and the Fermi pockets are located near the $\Gamma$ point ($\bs{k}=0$), where Kramers Weyl fermions can be expected to reside by the general symmetry arguments outlined above.
We show, however, that an additional set of Fermi pockets is relevant for the low-energy electronic structure of $\beta $-Ag$_2$Se as well, which originates from spinless Weyl nodes that are \textit{annihilated} by SOC -- a curious contrast to the Kramers Weyl nodes \textit{created} by SOC.
Our magnetotransport measurements reach the quantum limit (QL) of these Fermi pockets with a magnetic field as low as 3.2~T. In magnetic fields below the QL, we observed giant Shubnikov-de Haas (SdH) oscillations, on which a non-trivial Berry curvature contributed from the annihilated Weyl fermions is imprinted. Furthermore, we observed a negative longitudinal magnetoresistance (LMR) in a magnetic field beyond the QL. $\beta $-Ag$_2$Se presents  a rare example of topological semimetals in which a negative LMR is concomitant with the QL.

\section{Symmetry and Electronic structure}

Before reporting the experimental results in detail, we discuss the symmetries and electronic structure of $\beta $-Ag$_2$Se. Naumannite crystallizes in the orthorhombic space
group 19 ($P2_12_12_1$) with two crystallographically distinct silver atoms and one selenium atom \cite{ag2seold, ag2sezaac2008} (see upper inset of Fig.~\ref{Fig1}).
Space group 19 has very low symmetry: besides time-reversal, it features only three $C_2$ screw rotations around the principal axes: $C_2^x$, $C_2^y$, $C_2^z$.

The low-energy band structure of $\beta $-Ag$_2$Se arises entirely from a pair of electron and hole-like bands around the $\Gamma$ point.~\cite{ag2te2prl2011fang} If SOC is neglected [see Fig.~\ref{Fig6}(b)], these bands are doubly degenerate due to spin and carry opposite eigenvalues under the spinless $C_2^x$ and $C_2^y$ rotations. As a consequence, they cannot hybridize along the $\Gamma$-X and $\Gamma$-Y lines in the Brillouin zone. Rather, they show a pair of linear crossings along each of these lines. These linear crossings are four spinless Weyl cones, where those on the $\Gamma$-X line carry opposite chiral charge from those on the $\Gamma$-Y line. When SOC is included perturbatively, these Weyl degeneracies cannot be lifted due to their chiral charge. Rather, each spinless Weyl node splits into a pair of spinful Weyl nodes that is pinned to the $k_x$-$k_y$ plane. As SOC increases to the estimated physical value in $\beta $-Ag$_2$Se, spinful Weyl nodes that originated from different spinless Weyl nodes annihilate in the $k_x$-$k_y$ plane, leaving behind an electron-like Fermi pocket along the $\Gamma$-X line [Fig.~\ref{Fig6}(c)].

Alongside annihilating the band-inversion Weyl nodes, SOC also creates Kramers Weyl nodes located at the $\Gamma$ point. As a consequence, the hole-like Fermi surface centered at $\Gamma$ is split into two sheets [Fig.~\ref{Fig6}(c)], each of which is fully spin polarized with a hedgehog spin structure that is pointing toward the Weyl node on one sheet and away from the Weyl node on the other sheet.
The hedgehog spin structure is a direct consequence of the topological character of the Weyl node, which is a monopole of Berry curvature in momentum space.
The presence of two Fermi pockets in this Kramers Weyl semimetal, where one encloses the other, obscures some characteristics typically associated with condensed matter realizations of Weyl fermions. For example, no Fermi-arc states~\cite{chang2016kramers} would appear on the surface of $\beta $-Ag$_2$Se. On the other hand, Kramers Weyl systems uniquely allow to observe other characteristics of Weyl fermions, not only because of their enhanced symmetry and generic appearance, but also since the Weyl nodes are isolated in energy-momentum space.

A more detailed discussion of an effective $\bs{k}\cdot\bs{p}$ model for the Kramers Weyl node and the spinless Weyl nodes close to the Fermi energy, as well as the symmetry arguments outlined above can be found in Appendix.

\section{Crystal Growth, Characterization and Experimental Methods}
\begin{figure}[t]
\centering
  \includegraphics[clip, width=0.45\textwidth]{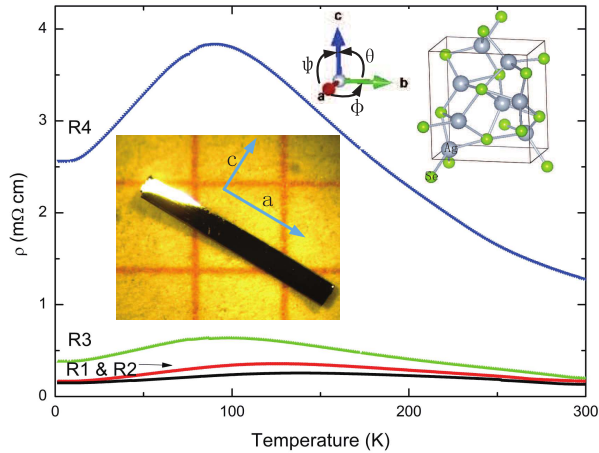}\\
  \caption{ Temperature dependent resistivity of four representative $\beta $-Ag$_2$Se single crystals. The resistivity along the $\bf{a}$ direction is similar to the previously reported values in polycrystalline samples with comparable carrier density $n$ \cite{xu_large_1997}. Upper inset: unit cell and definition of angles $\theta $, $\phi $ and $\psi$ with respect to the three principal axes. Central inset: a microscopical image of $\beta $-Ag$_2$Se crystal, with scale $1\times 1$ mm in the background.}
  \label{Fig1}
\end{figure}

In contrast to ample studies on polycrystalline $\beta $-Ag$_2$Se \cite{ag2sepressure2014, nanoag2sejacs2001, Yu_Ag2Se}, there is no report until now on macro-size single crystals, to the best of our knowledge. Its first order, polymorphic structural transition at 406~K hinders large single crystal formation because the cubic $\alpha $-phase prones to metamorphose to a multi-domain polycrystalline $\beta $-phase during cooling \cite{ag2severyold, Kumashiro1996761}. In order to avoid this roadblock of the structural transition, we design a single crystal growth of $\beta $-Ag$_2$Se above 406 K via a modified self-vapor transfer method \cite{Szczerbakow200581}.


Polycrystalline Ag$_2$Se was ground and sealed in a long fused silica ampoule in vacuum which was then placed in a large temperature gradient from 773 K to near room temperature in a tube furnace for a week. Single crystals of $\beta $-Ag$_2$Se were found near the hot zone where the temperature was presumably about 500 -- 600~K during the growth. The yielded crystals have a ribbon-like shape with a flat hexagonal cross section (central inset of Fig.~\ref{Fig1}). A scrutiny under microscope found that the crystals indeed burgeoned from small seeds of $\alpha $-phase crystals attached on the inner surface of the ampoule. Single crystal X-ray diffraction confirmed the crystallographic parameters consistent with previous reports on micro-size crystals (Table \ref{tab1} and \ref{tab2}). The diffraction also revealed that the long axis of the ribbon is pointing along the crystallographic $\bf{a}$ direction, and the large surface is perpendicular to the $\bf{b}$ direction. Although the structural rearrangement during the $\alpha$ to $\beta$ phase transition is not ascertained yet \cite{ag2sezaac2008}, this crystal growth indicates that the $\beta $-phase is able to solidify from saturated vapor above the transition temperature.


These macro-size single crystals of $\beta $-Ag$_2$Se allow for a study of the electronic states via transport measurements with less extent of influence of the scattering from defects, particularly avoiding the grain boundary scattering in polycrystals.
In this paper, we focus on the resistivity measurements when the current $I$ is along the crystallographic $\bf{a}$ direction while the magnetic fields are pointing along different directions. (See the upper inset of Fig.~\ref{Fig1} for the parametrization of the magnetic field direction in terms of angles relative to the crystallographic directions.) All physical property characterizations in low fields were performed in a Quantum Design Physical Property measurement system (PPMS-9), adopting the four-wire method.

\begin{table}[tbp]
\caption{
Crystallographic Data for $\beta $-Ag$_2$Se
}
\label{tab1}
\begin{ruledtabular}
\begin{tabular}{ll}
 Space group & $P2_12_12_1$\\
$a$({\AA}) & $4.3503(5)$\\
$b$({\AA}) & $7.0434(10)$\\
$c$({\AA}) & $7.6779(10)$\\
$V$({\AA}$^3$) & $235.26(5)$\\
$Z$ & $4$\\
$T$(K) of data collection & $180$\\
Crystal size (mm) & $0.13\times 0.06\times 0.04$\\
Radiation & Mo$K_\alpha $~$0.7107${\AA}\\
Collection region & $-3\leq h\leq 5$\\
& $-8\leq k\leq 8$\\
& $-9\leq l\leq 7$\\
$2\theta $ limit & $3.93^{\circ }\leq 2\theta \leq 25.98^{\circ }$\\
No. of measured reflections & $461$\\
No. of variable parameters & $29$\\
$R$, $R_w$ & $0.0626$, $0.1503$\\
\end{tabular}
\end{ruledtabular}
\end{table}

\begin{table}[tbp]
\caption{
Atomic Parameters and Anisotropic Thermal Vibration Parameters for $\beta $-Ag$_2$Se
}
\label{tab2}
\begin{ruledtabular}
\begin{tabular}{lcccc}
Atom & $x$ & $y$ & $z$ & $U_{iso}$ \\
\hline
Ag1 & $0.1525(4)$ & $0.3839(3)$ & $0.9523(3)$ & $0.0197(6)$ \\
Ag2 & $0.4769(4)$ & $0.2266(3)$ & $0.6357(3)$ & $0.0221(6)$ \\
Se & $0.1145(5)$ & $0.0034(3)$ & $0.8445(3)$ & $0.0116(6)$ \\
\end{tabular}

\begin{tabular}{lcccccc}
\multicolumn{7}{c}{$U_{ij}\times 100$}\\
Atom & $U_{11}$ & $U_{22}$ & $U_{33}$ & $U_{12}$ & $U_{13}$ & $U_{23}$ \\
\hline
Ag1 & $2.25(10)$ & $1.62(11)$ & $2.03(11)$ & $-0.33(9)$ & $-0.46(8)$ & $-0.04(8)$\\
Ag2 & $2.63(11)$ & $1.49(11)$ & $2.50(11)$ & $-0.40(8)$ & $0.42(7)$ & $-0.63(7)$\\
Se & $1.64(10)$ & $0.76(12)$ & $1.07(11)$ & $0.0$ & $ 0.0$ & $0.0$\\
\end{tabular}
\end{ruledtabular}
\end{table}

\section{Experiment and Data analysis}
\subsection{Magnetoresistance at different temperatures}

\begin{figure}[t]
\centering
  \includegraphics[clip, width=0.48\textwidth]{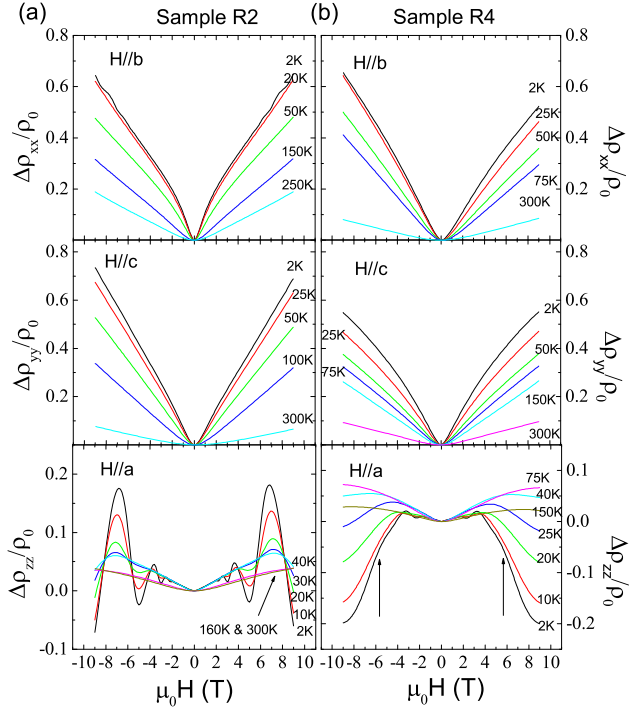}\\
  \caption{ MR ($\Delta \rho /\rho _0 $) for samples R2 (a) and R4 (b) along three principal axes at representative temperatures. From top to bottom: $\Delta \rho _{xx}$,  $\Delta \rho _{yy}$ and  $\Delta \rho _{zz}$ are for $H\parallel \bf{b}$, $\bf{c}$ and $\bf{a}$, respectively. The arrows in the lower right panel show the shoulder-like anomaly above the QL (see more details in text).  }
  \label{Fig2}
\end{figure}

The profile of the temperature dependent resistivity $\rho(T)$ for the single crystals of $\beta$-Ag$_2$Se is characterized by a broad hump at around 100~K (Fig.~\ref{Fig1}). This hump-like feature in $\rho(T)$ was also observed in polycrystals previously \cite{xu_large_1997, husmann_megagauss_2002}. Comparing the four representative samples labeled as R1--R4, we found that the resistivity of samples R1 and R2 from one growth batch is an order of magnitude smaller than that of samples R3 and R4 from a second growth batch.
Hall measurements ($I\parallel \bf{a}$, $H\parallel \bf{b}$) reveal that all samples are n-type for the whole temperature range and the carrier density is $n=4.9\times 10^{19}\mathrm{cm}^{-3} $ and  $2.4\times 10^{18}\mathrm{cm}^{-3} $ for samples R1 and R4 at 2~K, respectively (see Table~\ref{tab}).
The n-type carriers likely come from a small amount of deficiencies of selenium atoms which are commonly observed in other selenide compounds \cite{jia_BTS_2011}.
Both R1 and R4 show a mobility $\mu = 1/\rho ne $ of about 1000 cm$^2/\mathrm{Vs}$ at 2~K, leading to $\omega _{\mathrm{C}}\tau =\mu B<1$ at 9~T, where $\omega _{\mathrm{C}}$ is the cyclotron frequency and $\tau$ the scattering time.

Figure~\ref{Fig2}~(a) and~(b) show the magnetoresistance (MR = $\Delta \rho _H/\rho _0$) of the samples R2 and R4 along the three principal axes at different temperatures.
The transverse MR for $H\parallel \bf{b}$ and $\bf{c}$ for both samples shows a similar profile at different temperatures: the MR crosses over from a quadratic dependence on $H$ at low field to a linear and unsaturated increase up to 9~T.
Weak SdH oscillations occur on the linear background when $T<10$~K and $\mu _0H>5$ T for $H\parallel \bf{b}$ in R1 and R2 (the samples with larger $n$), but are absent in R3 and R4 (the samples with lower $n$).

At first glance, the profile of LMR for $H\parallel \bf{a}$ for R2 seems to be sheerly different from that for R4 at low temperatures. The LMR for R2 shows strong SdH oscillations up to 9 T, while that for R4 shows weak oscillations in low field, and then plummets to negative in a magnetic field stronger than 3.2 T. At higher temperatures, the negative LMR survives up to 25~K while the oscillations fade out at 10 K. The data of R2 at different temperatures also has an LMR with SdH oscillations and partially negative shape in high field.
Even at 40 K, when the oscillations have faded out completely, the LMR still bends down at 7 T. All of the interesting features can only be observed at low temperatures: the LMR for both samples degenerate to a small and positive signal when $T>100$~K.

An analysis of the SdH oscillations for $H\parallel \bf{a}$ for different samples sheds light on the ostensible difference of their LMR. We employed the Lifshitz-Onsager rule for the quantization of the Fermi surface cross-sectional area $S_{\mathrm{F}}$ in R2 and R4 as a function of magnetic field $B$
\begin{equation}
S_{\mathrm{F}} \frac{\hbar }{eB}=2\pi (N+\gamma ),
\label{eqn1}
\end{equation}
where $\hbar$ is the reduced  Planck constant, $e$ is the elementary charge, $N$ is the Landau level (LL) index, and $\gamma $ is the Onsager phase.
The SdH oscillations of R2 originate from a small electron pocket with $S_{\mathrm{F},\bf{a}}=8.1$~T, and the peaks at $\pm 6.8$~T correspond to $N$ = 1.
For R4, the SdH oscillations originate from an even smaller electron pocket with $S_{\mathrm{F},\bf{a}}= 3.8$~T, and a field stronger than 3.2~T drives the system beyond the QL.

Tracing the Landau indices of the SdH oscillations, we find that the negative part of LMR exactly occurs as long as the system is beyond the QL. From this point of view, the LMR of R2 and R4 indeed behaves similarly if we scale them with respect to the corresponding $S_{\mathrm{F}}$.
We can also conclude that the negative LMR is not a part of SdH oscillations for all the samples because it survives at higher temperatures than that the oscillations are washed out at.
Above the QL, the LMR of samples R3 and R4 shows a shoulder-like anomaly around 6~T (indicated by the arrows in Fig.~\ref{Fig2}).
This anomaly is always pegged at the fields close to $N$ = 1/2 when it is translated to LL indices for various samples with different $S_{\mathrm{F}}$.
The nature of this anomaly is not clear at this point.

\begin{figure}[t]
\centering
  \includegraphics[clip, width=0.45\textwidth]{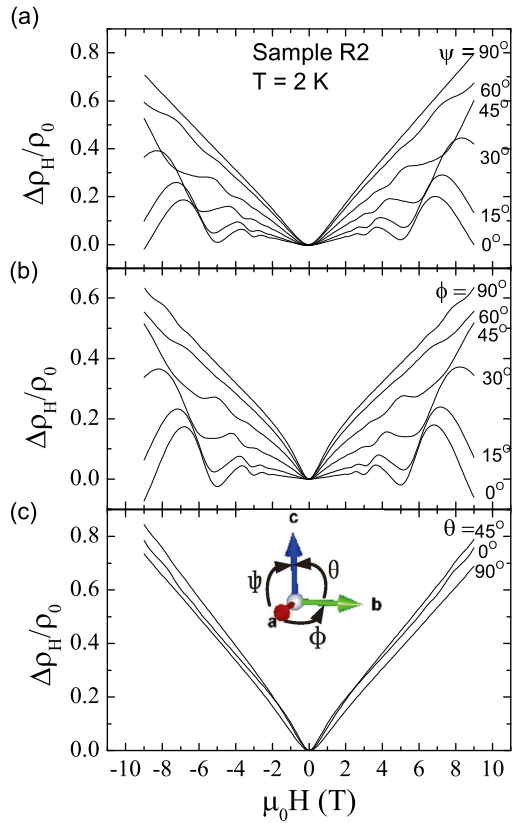}\\
 \caption{ Anisotropic MR for sample R2 when the direction of the field changes among three principal axes at 2 K.  No symmetrization between $\pm H$ was applied.   }
  \label{Fig3}
\end{figure}

\begin{figure}[t]
\centering
  \includegraphics[clip, width=0.5\textwidth]{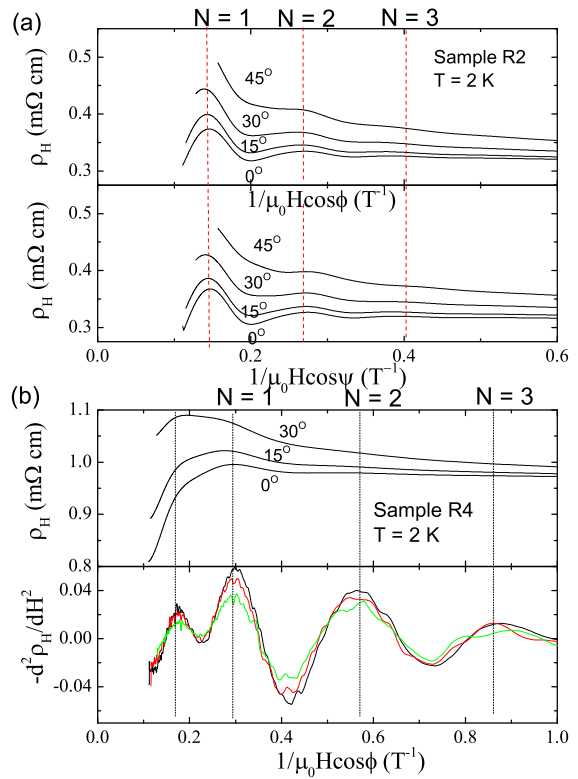}\\
 \caption{(a) The resistivity versus $1/\mu _0H_a = 1/(\mu _0H$cos$\phi )$ and $1/(\mu _0H$cos$\psi )$ for R2 at 2~K. (b)  The resistivity and its second derivative versus $1/\mu _0H_a = 1/(\mu _0H$cos$\phi )$ for R4 at 2~K. The shoulder-like anomaly at $N$ = 1/2 is clearly resolved as a peak in the second derivative. }
  \label{Fig3a}
\end{figure}

\subsection{Magnetoresistance at different angles}
In order to map out the electron pockets contributing to the SdH oscillations, we measured the MR for R2 and R4 while the direction of the magnetic field pointed along along angles deviating from the three principle axes at 2 K.  The MR changes are small when the direction of field is tilted from along $\bf{b}$ to along $\bf{c}$. On the other hand, we observed similar patterns of MR changes when the field is tilted from along $\bf{a}$ to along $\bf{b}$ and $\bf{c}$. Strong SdH oscillations in sample R2 keep pronounced for $\phi $ and $\psi < 60^\circ $ while the negative part of MR is compensated by the linear part of the transversal MR when $\phi $ and $\psi > 30^\circ $ [Fig. \ref{Fig3}(a) and (b)].


An analysis of the resistivity with respect to the reciprocal of the field projection on the $\bf{a}$ axis is shown in Fig.~\ref{Fig3a}(a). A clear 1/cos$\phi $ and 1/cos$\psi $ dependence up to $60^\circ $ indicates that the frequencies merely depend on the field component along the $\bf{a}$ direction irrespective of whether the field was tilted towards $\bf{b}$ or $\bf{c}$. Such angular dependence is characteristic of an anisotropic ellipsoid-like Fermi pocket. The MR for R4 with lower oscillation frequency in tilted magnetic field changes in the same manner. This similar angular dependence indicates that the pockets in R4 and R2 are cognate, albeit with different chemical potentials.

\begin{figure}[t]
\centering
  \includegraphics[clip, width=0.5\textwidth]{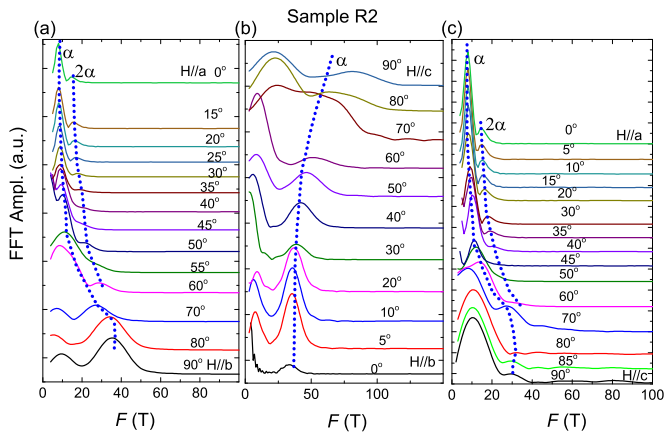}\\
  \caption{ FFT analysis of SdH oscillations at different angles. Dashed guide lines show the angular dependence of  frequency $\alpha$. These extra low-field frequency peaks in (b) are ill-shaped and angle-independent which indicates that they should not be associated with an actual closed cyclotron orbit in  $\beta$-Ag$_2$Se. }
  \label{Fig5}
\end{figure}

A fast Fourier transform (FFT) analysis for the field along different directions can trace subtle changes in the frequencies and map out the shape of the Fermi surface precisely.
Angle-dependent extremal orbits on Fermi surfaces are resolved as frequency peaks in the FFT in Fig.~\ref{Fig5}. Our analysis reveals that the large frequencies for $H\parallel \bf{b}$ and $\bf{c}$ indeed stem from the same electron pockets for $H\parallel \bf{a}$ with lower frequency. We denote this frequency, which maps out the anisotropic Fermi surface, by $\alpha$. This result is consistent with the angular dependent analyses presented above.

\begin{figure}[t]
\centering
  \includegraphics[clip, width=0.38\textwidth]{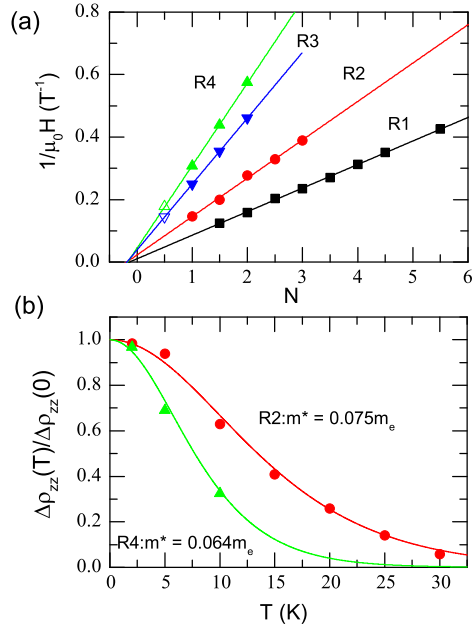}\\
  \caption{ SdH analyses for LMR for different samples.  (a) Landau level indices for all four samples. Two open symbols present the anomaly at $N$ = 1/2 for R3 and R4.  (b) Temperature dependence of the amplitude of the SdH oscillations at $N$ = 1 for R2 and R4.  }
  \label{Fig2a}
\end{figure}

\begin{table*}[tbp]
\caption{
Some electrical transport parameters for samples R1 to R4. $S_{\mathrm{F},\bf{a}}$, $k_{\mathrm{F},\bf{a}}$ and $S_{\mathrm{F},\bf{b}}$, $k_{\mathrm{F},\bf{b}}$ are determined from the SdH frequencies for the fields along $\bf{a}$ and $\bf{b}$ directions, respectively. $E_{\mathrm{F}}$ and $v_{\mathrm{F}}$ are obtained from the $T$ dependent amplitude of oscillations for the field along the $\bf{a}$ direction. Dash entries (N/A) signify that the respective quantities have not been measured or observed.
}
\label{tab}
\begin{ruledtabular}
\begin{tabular}{lccccccccc}
 & $n$ & $S_{\mathrm{F},\bf{a}}$ & $k_{\mathrm{F},\bf{a}}$ & $\gamma $ & m$^{\ast }$ & $E_{\mathrm{F}}$ & $v_{\mathrm{F}}$ & $S_{\mathrm{F},\bf{b}}$ & $k_{\mathrm{F},\bf{b}}$ \\
 & cm$^{-3}$ & T & {\AA}$^{-1}$ & & m$_e$ & meV & $10^5$ ms$^{-1}$ & T & {\AA}$^{-1}$ \\
\hline
R1 & $4.9\times 10^{19}$ & 13.3 & 0.02 & $-0.14\pm 0.04$ & 0.1 & 30 & 2.3 & 49 & 0.039 \\
R2 & -- & 8.1 & 0.016 & $-0.18\pm 0.07$ & 0.075 & 27 & 2.5 & 36 & 0.033\\
R3 & -- & 4.7 & 0.012 & $-0.17\pm 0.01$ & -- & -- & -- & N/A & N/A \\
R4 & $2.4\times 10^{18}$ & 3.8 & 0.011 & $-0.18\pm 0.01$ & 0.064 & 18 & 2.2 & N/A & N/A \\
\end{tabular}
\end{ruledtabular}
\end{table*}

\section{Discussion}
\subsection{Fermi surface}
To put the observations above in context, we return to the band-structure calculations presented in Fig.~\ref{Fig6}.
We estimated the Fermi level ($E_{\mathrm{F}}$) by comparing the calculated cross section of the $\alpha$ electron pockets with the measured frequencies for sample R2.
The Fermi level is placed at the dashed line in Fig.~\ref{Fig6}~(c), while the inferred Fermi surface is shown in Fig.~\ref{Fig6}~(d). Our samples of $\beta $-Ag$_2$Se host a pair of small electron-like kidney-shaped Fermi sheets originating from the gapped spinless Weyl nodes annihilated by SOC [marked in yellow in Fig.~\ref{Fig6}~(d)]. The anisotropy ratios of the calculated Fermi surface cross sections are 3.8 for $S_{\mathrm{F},\bf{b}}$/$S_{\mathrm{F},\bf{a}}$  and 4.0 for $S_{\mathrm{F},\bf{c}}$/$S_{\mathrm{F},\bf{a}}$. These ratios are consistent with the experimental results shown in Table~\ref{tab}. Besides the electron-like pockets, there are two large hole-like Fermi sheets enclosing the Kramers Weyl point at $\Gamma$ according to the band structure calculation [marked in green in Fig.~\ref{Fig6}~(d)]. However, these hole pockets do not show any discerning transport signatures in SdH and Hall measurements. The absence of the contribution from the predicted big hole pockets may be attributed to their low mobility and large effective mass.

Further information about the electron pockets from the spinless Weyl nodes that have been annihilated by SOC can be deduced from the change of the SdH oscillations at different temperatures \cite{moinmetals, rashbabook}:
\begin{equation}
\rho _H=\rho _0\left\{1+A(B,T)\cos\left[2\pi (S_{\mathrm{F}}/B+\gamma )\right]\right\},
\label{eqn2}
\end{equation}
where
\begin{equation}
A(B,T)\propto \mathrm{exp}\left(-\frac{2\pi ^2k_{\mathrm{B}}T_{\mathrm{D}}}{\hbar \omega_{\mathrm{C}}}\right)\  \frac{2\pi ^2k_{\mathrm{B}}T/\hbar \omega_{\mathrm{C}}}{\sinh(2\pi ^2k_{\mathrm{B}}T/\hbar \omega_{\mathrm{C}})}.
\label{eqn3}
\end{equation}
In Eq.~\eqref{eqn3}, $T_{\mathrm{D}}$ is the Dingle temperature, $k_{\mathrm{B}}$ is the Boltzmann constant, and the cyclotron frequency $\omega_{\mathrm{C}} = eB/m^*$ (where $m^*$ is the effective mass).
Based on Eq.~\eqref{eqn2}, the peak and valley positions of $\rho _{zz}$ are indexed as integers $N$ and half-integers, respectively (because 1/$\rho_{zz}$ = $\sigma _{zz}\propto1/\nu_{\mathrm{F}}$, where $\nu_{\mathrm{F}}$ is density of states at the Fermi level) \cite{WCMberryphase}.
Extrapolation of the lines of $N$ versus $1/\mu _0H$ for all four samples [Fig.~\ref{Fig2a}~(a)] leads to the intercepts at $1/\mu _0H=0$ clustered around $\gamma = -0.15\pm 0.05$, despite that $S_{\mathrm{F},\bf{a}}$ varies from 13.3~T to 3.8~T from sample to sample.
This near-to-zero phase shift, with zero being the alleged value for a 3D Dirac/Weyl semimetal \cite{WCMberryphase, Mikitik_berry_2012, Murakawascience2013}, may be interpreted as indication for a nontrivial Berry phase. Our anaylses indicate that the annihilation of the spinless Weyl cone does not nullify Berry curvature contributions.
The effective mass $m^*=eB/\omega _{\mathrm{C}} $ was obtained from the temperature dependence of the peak amplitude at $N$ = 1 for R2 and R4, yielding $0.075\,m_{\mathrm{e}}$ and $0.064\,m_{\mathrm{e}}$, respectively [see Fig.~\ref{Fig2a}~(b)].
If we assume a linear energy dispersion, we can estimate the Fermi wave vector $k_{\mathrm{F}}=\sqrt{2eS_{\mathrm{F}}/\hbar } $, the Fermi velocity $v_{\mathrm{F}}=\hbar k_{\mathrm{F}}/m^*$ and the Fermi energy $E_{\mathrm{F}}=v_{\mathrm{F}}^2m^* $ (listed in Table \ref{tab}). The mobility derived from the SdH oscillations is in the order of $10^3$ cm$^2$/Vs, comparable with that from the Hall measurements.

\begin{figure}[t]
  \begin{center}
  \includegraphics[clip, width=0.38\textwidth]{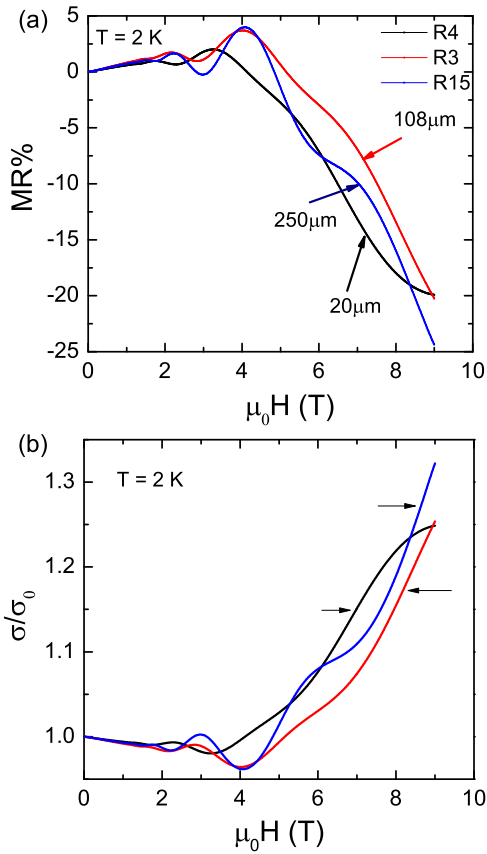}\\
 \caption{ (a) LMR for three samples of $\beta $-Ag$_2$Se with different thickness taken from the same growth batch. (b) Magneto-conductance for the three samples. The arrows indicate that a linear dependent magneto-conductance occurs when the field is higher than the position of the shoulder-like anomaly at $N$ = 1/2. }
  \label{Fig4}
  \end{center}
\end{figure}
\subsection{Negative longitudinal magnetoresistance }
There is a caveat in the magneto-transport measurements for bulk materials when the magnetic field and current are in parallel \cite{mrinmetals}: an inhomogeneous distribution of the electric current can induce a MR that is dramatically different from the actual  bulk MR of the system. The current jetting effect in strong magnetic fields when $\omega _{\mathrm{C}}\tau =\mu B\gg1$ can induce negative LMR in various materials including polycrystalline silver chalcogenide \cite{ag2seteprl2005, Jia}. Such spurious effect must be excluded before we claim any intrinsic negative LMR in single-crystalline $\beta$-Ag$_2$Se.

In order to clarify that the current jetting effect is minor and the observed negative LMR in our single crystalline $\beta$-Ag$_2$Se is intrinsic, we tested several samples with different thickness. Postulating that the negative LMR is merely due to a current distribution in thickness, the measured LMR should differ for these samples. In contrast, as shown in Fig~\ref{Fig4}. (a), three samples' LMR behave very similar while their thicknesses vary by one order of magnitude. Figure~\ref{Fig4}~(b) shows the magneto-conductance ($\sigma_{zz}$ = $1/\rho_{zz}$) for the three samples. We observe a linear field-dependence of $\sigma$ for fields above the shoulder-like anomaly in the QL.

After ruling out a current jetting effect, we try to fathom the origin of the negative LMR in single crystalline $\beta$-Ag$_2$Se. This concomitant negative part of LMR in the field-range above the QL should stem from the ultraquantum limit of the small electron-like pockets that originate from the spin-orbit annihilated Weyl cones. An Adler-Bell-Jackiw anomaly can cause a linear positive magneto-conductance in the QL for Weyl semimetals \cite{nielsen1983adler}. However a theoretical study taking into account the disorder scattering found that the linear magneto-conductance may also be caused by the interplay of the Landau degeneracy and scattering if $v_\mathrm{F}$ and $\tau$ depend on the field \cite{HZLlmr}. Besides topological semimetals, a topological trivial semiconductor may also bear a $H^2$ dependent magneto-conductance due to the ionized impurity scattering in a strong magnetic field \cite{Argyres_1956,Argyres_JPCS_1959, das3D}. This complexity in theory makes it difficult to ascribe the negative LMR in $\beta$-Ag$_2$Se to any cogent mechanism, based on our current observation. Nevertheless we comment that a negative LMR concomitant with the QL in $\beta$-Ag$_2$Se  was rarely reported for any semimetal before.


\subsection{Linear transversal magnetoresistance}
Previous studies showed that polycrystalline $\beta $-Ag$_2$Se manifests large, linear-field-dependent, transversal MR in both low and high temperature regimes \cite{xu_large_1997, husmann_megagauss_2002}.
This linear MR remains unsaturated as a function of magnetic field strength up to 50~T.
The origin of this unusual linear MR was interpreted as a quantum effect of electrons attributed to a linear dispersion at the conduction and valence band touching points (ie., a quantum MR) \cite{qlmrep2000abrikosov, qmrprb1998abrikosov}.  An alternative explanation is that spatial conductivity fluctuations induce a linear MR in these strongly inhomogenous polycrystals~\cite{parish_non-saturating_2003, hu_classical_2008}.
Our measurements on the single crystals may provide new evidence supporting the classical effect. Our band structure calculation and SdH oscillations reveal that the field in which the linear MR occurs is far less than the QL along the $\bf{b}$ and $\bf{c}$ directions which is the range of fields, in which the presumed quantum MR would be occurring \cite{qlmrep2000abrikosov, qmrprb1998abrikosov}. It is noteworthy that the linear MR of the polycrystals in the literature stretches to much lower field than what we observed for the single crystals. This difference can be well explained by the classical effect in which the stronger spatial conductivity fluctuations in polycrystals  lead to more palpable linear MR in lower field.

\section{Conclusion}
In summary, we synthesized macro-size single crystals of the Kramers Weyl semimetal candidate $\beta$-Ag$_2$Se and determined its fermiology in detail from SdH oscillations. The dominant SdH signal stems from small non-spin-degenerate and highly anisotropic Fermi surfaces of the bottom of a gapped spinless Weyl cone. As such, these electron pockets inherit a nonvanishing Berry curvature verified as a unified, nontrivial Onsager phase among several samples. A negative LMR concomitant with the QL of these small electron pockets was observed in strong magnetic field. Finally, we infer that the vapour transfer growth method bears  great potential for  synthesizing single crystals of new materials, especially for the thermally unstable compounds.

\begin{acknowledgments}
The authors thank Jun Xiong, Chen Fang and Fa Wang for helpful discussion, and Kaihui Liu for assistance with TEM measurements. The authors thank Jian Wang and Yuan Li for using their instruments. S. J. is supported by National Basic Research Program of China (Grant No. 2014CB239302) and National Natural Science Foundation of China (Grant No. 11774007). F.S. acknowledges support from the Swiss National Science Foundation. C.L. Zhang treats his research work as precious homework from Shui-Fen Fan.
\end{acknowledgments}
\bibliographystyle{unsrt}

\begin{thebibliography}{10}

\bibitem{Weyl_wanxiangang}
Xiangang Wan, Ari~M. Turner, Ashvin Vishwanath, and Sergey~Y. Savrasov.
\newblock Topological semimetal and fermi-arc surface states in the electronic
  structure of pyrochlore iridates.
\newblock {\em Phys. Rev. B}, 83:205101, May 2011.

\bibitem{TaAs_Chenyulin}
LX~Yang, ZK~Liu, Yan Sun, Han Peng, HF~Yang, Teng Zhang, Bo~Zhou, Yi~Zhang,
  YF~Guo, Marein Rahn, et~al.
\newblock Weyl semimetal phase in the non-centrosymmetric compound taas.
\newblock {\em Nature physics}, 11(9):728--732, 2015.

\bibitem{TaAs_theory_Hsin}
Shin-Ming Huang, Su-Yang Xu, Ilya Belopolski, Chi-Cheng Lee, Guoqing Chang,
  BaoKai Wang, Nasser Alidoust, Guang Bian, Madhab Neupane, Chenglong Zhang,
  et~al.
\newblock A weyl fermion semimetal with surface fermi arcs in the transition
  metal monopnictide taas class.
\newblock {\em Nature Communications}, 6, 2015.

\bibitem{TaAs_theory_Xidai}
Hongming Weng, Chen Fang, Zhong Fang, B.~Andrei Bernevig, and Xi~Dai.
\newblock Weyl semimetal phase in noncentrosymmetric transition-metal
  monophosphides.
\newblock {\em Phys. Rev. X}, 5:011029, Mar 2015.

\bibitem{TaAs_Arpes_Hongding}
B.~Q. Lv, H.~M. Weng, B.~B. Fu, X.~P. Wang, H.~Miao, J.~Ma, P.~Richard, X.~C.
  Huang, L.~X. Zhao, G.~F. Chen, Z.~Fang, X.~Dai, T.~Qian, and H.~Ding.
\newblock Experimental discovery of weyl semimetal taas.
\newblock {\em Phys. Rev. X}, 5:031013, Jul 2015.

\bibitem{TaAs_Arpes_Hasan}
Su-Yang Xu, Ilya Belopolski, Nasser Alidoust, Madhab Neupane, Guang Bian,
  Chenglong Zhang, Raman Sankar, Guoqing Chang, Zhujun Yuan, Chi-Cheng Lee,
  Shin-Ming Huang, Hao Zheng, Jie Ma, Daniel~S. Sanchez, BaoKai Wang, Arun
  Bansil, Fangcheng Chou, Pavel~P. Shibayev, Hsin Lin, Shuang Jia, and M.~Zahid
  Hasan.
\newblock Discovery of a weyl fermion semimetal and topological fermi arcs.
\newblock {\em Science}, 2015.

\bibitem{nodal_burkov}
A.~A. Burkov, M.~D. Hook, and Leon Balents.
\newblock Topological nodal semimetals.
\newblock {\em Phys. Rev. B}, 84:235126, Dec 2011.

\bibitem{bian2016topological}
Guang Bian, Tay-Rong Chang, Raman Sankar, Su-Yang Xu, Hao Zheng, Titus Neupert,
  Ching-Kai Chiu, Shin-Ming Huang, Guoqing Chang, Ilya Belopolski, et~al.
\newblock Topological nodal-line fermions in spin-orbit metal pbtase2.
\newblock {\em Nature communications}, 7, 2016.

\bibitem{schoop2016dirac}
Leslie~M Schoop, Mazhar~N Ali, Carola Stra{\ss}er, Andreas Topp, Andrei
  Varykhalov, Dmitry Marchenko, Viola Duppel, Stuart~SP Parkin, Bettina~V
  Lotsch, and Christian~R Ast.
\newblock Dirac cone protected by non-symmorphic symmetry and three-dimensional
  dirac line node in zrsis.
\newblock {\em Nature communications}, 7, 2016.

\bibitem{Madhab_ZrSiS}
Madhab Neupane, Ilya Belopolski, M.~Mofazzel Hosen, Daniel~S. Sanchez, Raman
  Sankar, Maria Szlawska, Su-Yang Xu, Klauss Dimitri, Nagendra Dhakal, Pablo
  Maldonado, Peter~M. Oppeneer, Dariusz Kaczorowski, Fangcheng Chou, M.~Zahid
  Hasan, and Tomasz Durakiewicz.
\newblock Observation of topological nodal fermion semimetal phase in zrsis.
\newblock {\em Phys. Rev. B}, 93:201104, May 2016.

\bibitem{pts_ZCL}
Cheng-Long Zhang, Zhujun Yuan, Guang Bian, Su-Yang Xu, Xiao Zhang, M.~Zahid
  Hasan, and Shuang Jia.
\newblock Superconducting properties in single crystals of the topological
  nodal semimetal ${\mathrm{pbtase}}_{2}$.
\newblock {\em Phys. Rev. B}, 93:054520, Feb 2016.

\bibitem{bradlyn2016beyond}
Barry Bradlyn, Jennifer Cano, Zhijun Wang, MG~Vergniory, C~Felser, RJ~Cava, and
  B~Andrei Bernevig.
\newblock Beyond dirac and weyl fermions: Unconventional quasiparticles in
  conventional crystals.
\newblock {\em Science}, 353(6299):aaf5037, 2016.

\bibitem{soluyanov2015type}
Alexey~A Soluyanov, Dominik Gresch, Zhijun Wang, QuanSheng Wu, Matthias Troyer,
  Xi~Dai, and B~Andrei Bernevig.
\newblock Type-ii weyl semimetals.
\newblock {\em Nature}, 527(7579):495--498, 2015.

\bibitem{wang2016hourglass}
Zhijun Wang, Aris Alexandradinata, Robert~J Cava, and B~Andrei Bernevig.
\newblock Hourglass fermions.
\newblock {\em Nature}, 532(7598):189--194, 2016.

\bibitem{chang2016kramers}
Guoqing Chang, Daniel~S Sanchez, Benjamin~J Wieder, Su-Yang Xu, Frank
  Schindler, Ilya Belopolski, Shin-Ming Huang, Bahadur Singh, Di~Wu, Titus
  Neupert, et~al.
\newblock Kramers theorem-enforced weyl fermions: Theory and materials
  predictions.
\newblock {\em arXiv preprint 1611.07925}, 2016.



\bibitem{ag2seold}
G.~A. Wiegers.
\newblock The crystal structure of the low-temperature form of silver selenide.
\newblock {\em American mineralogist}, 56:1882, 1971.

\bibitem{ag2sezaac2008}
Heinrich Billetter and Uwe Ruschewitz.
\newblock Structural phase transitions in $\mathrm{Ag_2Se}$ (naumannite).
\newblock {\em Zeitschrift f¨¹r anorganische und allgemeine Chemie},
  634(2):241--246, 2008.

\bibitem{ag2te2prl2011fang}
Wei Zhang, Rui Yu, Wanxiang Feng, Yugui Yao, Hongming Weng, Xi~Dai, and Zhong
  Fang.
\newblock Topological aspect and quantum magnetoresistance of
  $\beta-\mathrm{Ag_2Te}$.
\newblock {\em Phys. Rev. Lett.}, 106:156808, Apr 2011.

\bibitem{ag2sepressure2014}
Zhao Zhao, Shibing Wang, Artem~R. Oganov, Pengcheng Chen, Zhenxian Liu, and
  Wendy~L. Mao.
\newblock Tuning the crystal structure and electronic states of
  $\mathrm{Ag_2Se}$: Structural transitions and metallization under pressure.
\newblock {\em Phys. Rev. B}, 89:180102, May 2014.

\bibitem{nanoag2sejacs2001}
Byron Gates, Yiying Wu, Yadong Yin, Peidong Yang, and Younan Xia.
\newblock Single-crystalline nanowires of $\mathrm{Ag_2Se}$ can be synthesized
  by templating against nanowires of trigonal $\mathrm{Se}$.
\newblock {\em Journal of the American Chemical Society}, 123(46):11500--11501,
  2001.

\bibitem{Yu_Ag2Se}
Jaemin Yu and Hoseop Yun.
\newblock {Reinvestigation of the low-temperature form of Ag${\sb 2}$Se
  (naumannite) based on single-crystal data}.
\newblock {\em Acta Crystallographica Section E}, 67(9):i45, Sep 2011.


\bibitem{ag2severyold}
P.~Rahlfs.
\newblock The cubic high temperature modificators of sulfides, selenides and
  tellurides of silver and of uni-valent copper.
\newblock {\em Z. Phys. Chem}, B 31:157, 1936.

\bibitem{Kumashiro1996761}
Y.~Kumashiro, T.~Ohachi, and I.~Taniguchi.
\newblock Phase transition and cluster formation in silver selenide.
\newblock {\em Solid State Ionics}, 86¨C88, Part 2(0):761 -- 766, 1996.
\newblock Proceedings of the 10th International Conference on Solid State
  Ionics.

\bibitem{Szczerbakow200581}
Andrzej Szczerbakow and Ken Durose.
\newblock Self-selecting vapour growth of bulk crystals - principles and
  applicability.
\newblock {\em Progress in Crystal Growth and Characterization of Materials},
  51(1¨C3):81 -- 108, 2005.


\bibitem{xu_large_1997}
R.~Xu, A.~Husmann, T.~F. Rosenbaum, M.-L. Saboungi, J.~E. Enderby, and P.~B.
  Littlewood.
\newblock Large magnetoresistance in non-magnetic silver chalcogenides.
\newblock {\em Nature}, 390(6655):57--60, nov 1997.


\bibitem{husmann_megagauss_2002}
A.~Husmann, J.~B. Betts, G.~S. Boebinger, A.~Migliori, T.~F. Rosenbaum, and
  M.-L. Saboungi.
\newblock Megagauss sensors.
\newblock {\em Nature}, 417(6887):421--424, may 2002.

\bibitem{jia_BTS_2011}
Shuang Jia, Huiwen Ji, E.~Climent-Pascual, M.~K. Fuccillo, M.~E. Charles, Jun
  Xiong, N.~P. Ong, and R.~J. Cava.
\newblock Low-carrier-concentration crystals of the topological insulator
  bi${}_{2}$te${}_{2}$se.
\newblock {\em Phys. Rev. B}, 84:235206, Dec 2011.






\bibitem{moinmetals}
D.~Shoenberg.
\newblock {\em Magnetic oscillations in metal}.
\newblock Cambridge University Press, 2009.


\bibitem{rashbabook}
G.~Landwehr and E.~I. Rashba.
\newblock {\em Landau Level spectroscopy}, volume 27.2.
\newblock North-Holland, Amsterdam, 1991.

\bibitem{WCMberryphase}
C.M. Wang, Hai-Zhou Lu, and Shun-Qing Shen.
\newblock Anomalous Phase Shift of Quantum Oscillations in 3D Topological Semimetals.
\newblock {\em Phys. Rev. Lett.}, 117:077201, Aug 2016.




\bibitem{Mikitik_berry_2012}
G.~P. Mikitik and Yu.~V. Sharlai.
\newblock Berry phase and the phase of the Shubnikov de Haas
  oscillations in three-dimensional topological insulators.
\newblock {\em Phys. Rev. B}, 85:033301, Jan 2012.

\bibitem{Murakawascience2013}
H.~Murakawa, M.~S. Bahramy, M.~Tokunaga, Y.~Kohama, C.~Bell, Y.~Kaneko,
  N.~Nagaosa, H.~Y. Hwang, and Y.~Tokura.
\newblock Detection of berry¡¯s phase in a bulk rashba semiconductor.
\newblock {\em Science}, 342(6165):1490--1493, 2013.





\bibitem{mrinmetals}
Alfred~Brian Pippard.
\newblock {\em Magnetoresistance in Metals}.
\newblock Cambridge University Press, 2009.

\bibitem{ag2seteprl2005}
Jingshi Hu, T.~F. Rosenbaum, and J.~B. Betts.
\newblock Current jets, disorder, and linear magnetoresistance in the silver
  chalcogenides.
\newblock {\em Phys. Rev. Lett.}, 95:186603, Oct 2005.

\bibitem{Jia}
Xi-Tong Xu and Shuang Jia.
\newblock Recent observations of negative longitudinal magnetoresistance in semimetal.
\newblock {\em Chinese Physics B}, 25(11):117204,2016.


\bibitem{nielsen1983adler}
Holger~Bech Nielsen and Masao Ninomiya.
\newblock The {Adler-Bell-Jackiw} anomaly and {Weyl} fermions in a crystal.
\newblock {\em Physics Letters B}, 130(6):389--396, 1983.

\bibitem{HZLlmr}
H.-Z. Lu and S.-Q Shen.
\newblock Quantum transport in topological semimetals under magnetic fields.
\newblock {\em Frontiers of Physics}, 12: 127201, 2017.

\bibitem{Argyres_1956}
P. N. Argyres, E. N. Adams.
\newblock Longitudinal magnetoresistance in the quantum limit.
\newblock {\em Phys. Rev.}, 104:900, 1956.

\bibitem{Argyres_JPCS_1959}
Petros~N. Argyres.
\newblock Galvanomagnetic effects in the quantum limit.
\newblock {\em Journal of Physics and Chemistry of Solids}, 8(0):124 -- 130, 1959.

\bibitem{das3D}
Pallab Goswami, J. H. Pixley, and S. Das Sarma.
\newblock Axial anomaly and longitudinal magnetoresistance of a generic three-dimensional metal.
\newblock {\em Phys. Rev. B}, 92, 075205, Aug 2015.



\bibitem{qmrprb1998abrikosov}
A.~A. Abrikosov.
\newblock Quantum magnetoresistance.
\newblock {\em Phys. Rev. B}, 58:2788--2794, Aug 1998.

\bibitem{qlmrep2000abrikosov}
A.~A. Abrikosov.
\newblock Quantum linear magnetoresistance.
\newblock {\em EPL (Europhysics Letters)}, 49(6):789, 2000.

\bibitem{parish_non-saturating_2003}
M.~M. Parish and P.~B. Littlewood.
\newblock Non-saturating magnetoresistance in heavily disordered
  semiconductors.
\newblock {\em Nature}, 426(6963):162--165, nov 2003.

\bibitem{hu_classical_2008}
Jingshi Hu and T.~F. Rosenbaum.
\newblock Classical and quantum routes to linear magnetoresistance.
\newblock {\em Nat Mater}, 7(9):697--700, sep 2008.










\end{thebibliography}

\clearpage

\section{Appendix: Effective $\bs{k}\cdot\bs{p}$ theory for spinless and Kramers Weyl points}
\label{app: k dot p}

We want to derive an effective model for the two pairs of bands near the Fermi energy of Ag$_2$Se that exist near the $\Gamma$-point of the Brillouin zone, where each pair is split by SOC.

Let us first look at the symmetries of the problem:
The space group P $2_1 2_1 2_1$ of Ag$_2$Se is generated by
\begin{equation}
\begin{split}
(x,y,z) \quad &\rightarrow \quad \left(\frac{1}{2} + x, \frac{1}{2} -y, -z\right), \\
&\rightarrow \quad \left(-x, \frac{1}{2} + y, \frac{1}{2}-z\right), \\
&\rightarrow \quad \left(\frac{1}{2} - x, -y, \frac{1}{2}+z\right).
\end{split}
\end{equation}

In the Bloch basis, the translation operators corresponding to the non-symmorphic part of the transformations are diagonal in the momenta $(k_x, k_y, k_z)^T = \bs{k}$, and therefore do not give constraints on the $\bs{k}\cdot\bs{p}$ Hamiltonian $H(\bs{k})$. We are then left with three independent $C_2$ rotations:
\begin{equation}
\begin{split}
&C_{2,x} \, \bs{k} \, C_{2,x}^{-1} \quad = \quad \left(k_x, -k_y, -k_z \right)^T, \\
&C_{2,y} \, \bs{k} \, C_{2,y}^{-1} \quad = \quad \left(-k_x, k_y, -k_z \right)^T, \\
&C_{2,z} \, \bs{k} \, C_{2,z}^{-1} \quad = \quad \left(-k_x, -k_y, k_z \right)^T.
\end{split}
\end{equation}

\subsection{The fate of spinless Weyl points}

We first consider the case without SOC, in which the two pairs of bands around $\Gamma$ remain spin-degenerate. From the first principles calculation, we know that they have the same $C_{2,z}$ eigenvalue along the $k_z$ axis and opposite eigenvalues under $C_{2,x}$ and $C_{2,y}$ along the $k_x$ and $k_y$ axis, respectively.
Henceforth, we can choose, without loss of generality, the representations
\begin{equation}
R_{2,x}=\tau_x,\qquad R_{2,y}=\tau_x, \qquad R_{2,z}=\openone_2,
\end{equation}
where $\tau_{x,y,z}$ are the three Pauli matrices and $\openone_2$ is the $2\times 2$ unit matrix. This spinless representation obeys $R_{2,x}R_{2,y}=R_{2,z}$, as it should .

Spinless time reversal symmetry is given by $T=K$, where $K$ is complex conjugation. Furthermore, we have $T \, \bs{k} \,  T^{-1} = - \bs{k}$, since flipping the direction of time reverses all velocities.
The most general spinless Hamiltonian that obeys both time-reversal and the three two-fold rotation symmetries is, to second order in $\bs{k}$, given by
\begin{equation}
H_{0}=(w_xk_x^2+w_yk_y^2+w_zk_z^2-m)\tau_x+a k_z \tau_y+b k_x k_y \tau_z.
\label{eq: spinless H0}
\end{equation}
Note that we have neglected the term proportional to the identity matrix, which is irrelevant for the discussion of symmetry or topology enforced band degeneracies. Hamiltonian~\eqref{eq: spinless H0} asymptotically describes an electron and a hole-like band. The Hamiltonian can feature up to four spinless Weyl points at the points in momentum space where each of the terms vanishes individually. For $w_x,w_y,w_z,m>0$, the first term vanishes on an ellipsoid in momentum space. The intersections of this ellipsoid with the $k_x$ and $k_y$ axis are the locations of the four Weyl points. This is in agreement with the observation from our DFT calculation, which exhibits these Weyl points if SOC is neglected.

We will now discuss how these Weyl points are perturbed if SOC is switched on.
It is clear that due to their chiral nature  infinitesimal perturbations cannot gap them out immeadiately, but rather split them in momentum space.
If the spin-space is acted on with the three Pauli matrices $\sigma_{x,y,z}$, the symmetries are now represented as follows. Time reversal is implemented by the operator $T = \mathit{K} \mathrm{i} \sigma_y$.
The spinful rotation operations are given by
\begin{equation}
R_{2,x}=\mathrm{i}\tau_x\otimes\sigma_x,\quad
R_{2,y}=\mathrm{i}\tau_x\otimes\sigma_y, \quad
R_{2,z}=\mathrm{i}\openone_2\otimes\sigma_z.
\end{equation}
We will only consider SOC terms up to linear order in $\bs{k}$. Obviously, there is no single SOC term that anticommutes with all terms forming a Weyl cone in Hamiltonian~\eqref{eq: spinless H0}. This is in accordance with our expectation that infinitesimal SOC cannot gap out the Weyl nodes.
The following SOC terms are symmetry allowed
\begin{equation}
\begin{split}
H_{\mathrm{SOC}}=&
\lambda_1\,\tau_y\otimes\sigma_z
+\lambda_2\,k_x\openone_2\otimes\sigma_x
+\lambda_3\,k_x\tau_x\otimes\sigma_x\\
&+\lambda_4\,k_x\tau_z\otimes\sigma_y
+\lambda_5\,k_y\openone_2\otimes\sigma_y
+\lambda_6\,k_y\tau_x\otimes\sigma_y\\
&+\lambda_7\,k_y\tau_z\otimes\sigma_x
+\lambda_8\,k_z\openone_2\otimes\sigma_z
+\lambda_9\,k_z\tau_x\otimes\sigma_3.
\end{split}
\end{equation}
Of these, the $\lambda_2$, $\lambda_5$, and $\lambda_8$ terms commute with the entire Hamiltonian~\eqref{eq: spinless H0} and do therefore \textit{per se} not lead to any change in the position of the Weyl nodes, but spin-split them in energy. The remaining terms, each on their own have the following impact:
The $\lambda_1$ term splits each of the four spinless Weyl nodes into two Weyl nodes separated in $k_z$ direction.
The $\lambda_3$ term splits each of the two spinless Weyl nodes on the $k_x$ axis into two Weyl nodes separated in $k_x$ direction.
The $\lambda_4$ term splits each of the two spinless Weyl nodes on the $k_x$ axis into two Weyl nodes separated in $k_y$ direction.
The $\lambda_6$ term splits each of the two spinless Weyl nodes on the $k_y$ axis into two Weyl nodes separated in $k_y$ direction.
The $\lambda_7$ term splits each of the two spinless Weyl nodes on the $k_y$ axis into two Weyl nodes separated in $k_x$ direction.
In summary, the SOC terms have the potential to split each of the spinless Weyl points into two Weyl nodes along the $k_x$, $k_y$, or $k_z$ direction. Time-reversal and rotation symmetries pin the Weyl nodes to lie in quadruplets on the high-symmetry planes $k_x$-$k_y$, $k_y$-$k_z$, or $k_x$-$k_z$.

As evidenced by our first-principles calculation, however, SOC in Ag$_2$Se is strong enough to gap out the Weyl nodes by annihilating them. As SOC is gradually increased, each spinless Weyl node splits into two Weyl nodes that separate in a way corresponding to dominant $\lambda_4$ and $\lambda_7$ terms in the $\bs{k}\cdot\bs{p}$ model. When SOC reaches the physical value it has in Ag$_2$Se, Weyl nodes that originated from different spinless Weyl cones have annihilated pairwise and there is a direct gap between the second and third band in all of momentum space.

\subsection{Emergence of Kramers Weyl points}

In the above we have largely neglected the discussion of the terms proportional to $\lambda_2$, $\lambda_5$, and $\lambda_8$ as they commute with the Hamiltonian~\eqref{eq: spinless H0}. They do, however, have an important effect on the spin-splitting of an individual band: Isolated Weyl nodes emerge in the form of Kramers degeneracies of \emph{every} pair of bands at the $\Gamma$ point in momentum space.

To show this within a $\bs{k}\cdot\bs{p}$ expansion, we focus on a single pair of bands only [as compared to two pairs of bands in Hamiltonian~\eqref{eq: spinless H0}].
The effective Hamiltonian $H(\bs{k})$ should be invariant under all $C_2$ rotations, as well as time reversal. As any two-dimensional matrix, we can write it as a linear combination of the identity and $\sigma$-matrices. In an expansion around the $\Gamma$-point in powers of the individual components of $\bs{k}$, time reversal symmetry excludes odd powers for the coefficient of $\openone_2$, and even powers for the coefficients of the $\sigma$-matrices. The $C_2$ rotation symmetries imply that every component of $\bs{k}$ has to be paired up with the same component of $\bs{\sigma}$, or with itself. To second order in $\bs{k}$, the Hamiltonian is then given by
\begin{equation}
\begin{split}
H(\bs{k}) = &v_x k_x \, \sigma_x + v_y k_y \, \sigma_y + v_z k_z \, \sigma_z \\
&+ (u_x k_x^2 + u_y k_y^2 + u_z k_z^2) \, \openone_2.
\end{split}
\label{eq: eff H Kramers Weyl}
\end{equation}
So just by symmetry, the band structure always contains a Weyl point at $\bs{k} = 0$.

This result is in qualitative agreement with our DFT calculation for Ag$_2$Se. We can determine the coefficients from the DFT calculation, for the lower pair of bands that cross the Fermi level, as
\begin{equation}
\begin{aligned}
&v_x = 0.079 \, \text{eV\AA}, \hspace{0.32cm} v_y = 0.066 \, \text{eV\AA}, \hspace{0.32cm}  v_z = 0.020 \, \text{eV\AA}, \\
&u_x= -14.8 \, \text{eV\AA$^2$}, u_y= -2.97 \, \text{eV\AA$^2$}, u_z= -1.55 \, \text{eV\AA$^2$}.
\end{aligned}
\label{eq: model parameters}
\end{equation}

The Chern numbers of the two Fermi surfaces are exactly opposite and there are no Fermi arcs visible on any surface.

\end{document}